\documentclass{article}

\usepackage{arxiv}

\usepackage[utf8]{inputenc} 
\usepackage[T1]{fontenc}    
\usepackage{hyperref}       
\usepackage{url}            
\usepackage{booktabs}       
\usepackage{amsfonts}       
\usepackage{nicefrac}       
\usepackage{microtype}      

\usepackage{multirow}
\usepackage{calc}
\usepackage{stmaryrd}
\usepackage{tikz}
\usetikzlibrary{shapes.symbols, patterns}
\usepackage{amsmath, amssymb}
\usepackage[utf8]{inputenc}
\usepackage{mwe}
\usepackage{graphicx}
\usetikzlibrary{shapes}
\usepackage{mathtools}
\usepackage{array}
\usepackage{algorithm}
\usepackage[noend]{algpseudocode}
\algnewcommand{\LineComment}[1]{\State \(\triangleright\) #1}
\DeclareMathOperator*{\argmin}{\arg\!\min}

\title{Short-term Wind Speed Forecasting based on LSSVM Optimized by Elitist QPSO}

\author{
  Ephrem Admasu Yekun \\
  School of Electrical and Computer Engineering\\
  Mekelle University\\
  Mekelle, Ethiopia\\
  \texttt{ephraim.admasu@gmail.com} \\
   \And
 Alem Haddush Fitwi \\
   School of Electrical and Computer Engineering\\
  Mekelle University\\
  Mekelle, Ethiopia\\
  \texttt{alemh29@gmail.com} \\
\And
 S. Karpaga Selvi \\
   School of Electrical and Computer Engineering\\
  Mekelle University\\
  Mekelle, Ethiopia\\
  \texttt{karpagaselvil@gmail.com}
  \And
 Anubhav Kumar \\
   School of Computing\\
  Mekelle University\\
  Mekelle, Ethiopia\\
  dr.anubhavkumar@gmail.com
}

\begin{document}
\maketitle

\begin{abstract}
Nowadays, wind power is considered as one of the most widely used renewable energy applications due to its efficient energy use and low pollution. In order to maintain high integration of wind power into the electricity market efficient models for wind speed forecasting are in high demand. The non-stationary and nonlinear characteristics of wind speed, however, makes the task of wind speed forecasting challenging. LSSVM has proven to be a good forecasting algorithm mainly for time series applications such as wind data. To boost the learning performance and generalization capablity of the algorithm, LSSVM has two hyperparameters, known as the regularization and kernel parameters, that require careful tuning. In this paper a modified QPSO algorithm is proposed that uses the principle of transposon operators to breed the personal best and global best particles of QPSO and improve global searching capabilities. The optimization algorithm is then used to generate optimum values for the LSSVM hyperparameters. Finally, the performance of the proposed model is compared with previously known
PSO and QPSO optimized LSSVM models. Empirical results show that the forecasting performance of proposed model is greatly improved when comapred to the competetive methods.

\end{abstract}


\section{Introduction}
The alarming oil scarcity and increasing global awareness about climate change is drawing people's attention to the exploitation of renewable energy technologies \cite{ahuja2016challenges}. One such energy, wind power, is regarded as one of the most attractive renewable energy sources due to its efficient energy use, environmental friendliness, and economical competetiveness. Further, wind turbines, which convert the kinetic  energy of wind into electrical energy offer electric power with lower installation costs, higher reliability, and cost-effective operation.

\begin{figure}
\centering
\includegraphics[scale=.45]{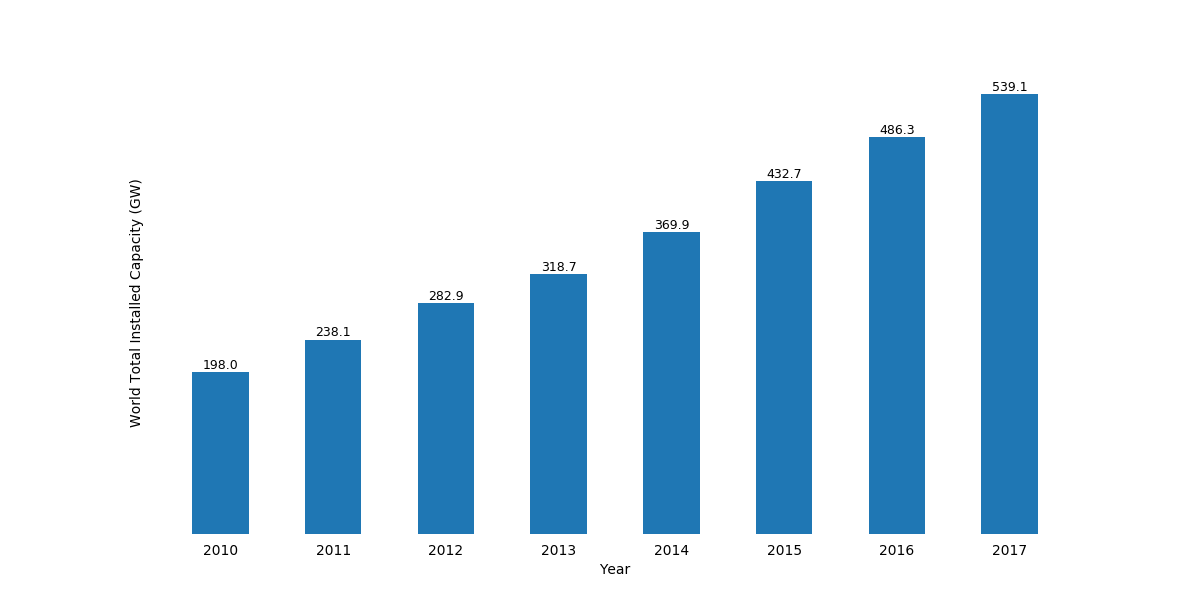}
\caption{\label{data} Global Cumulative Installed Capacity 2011--2017 \cite{globalwind2019}}
\end{figure}

The life span of wind turbines is 8 to 10 years and are easily decommissioned without causing much damage in the environment. These characteristics have algined wind speed in the direction of becoming the fastest-growing source of renewable energy \cite{sun2015wind}. The efforts to promote electrification using renewable energy are gaining momentum and large-scale wind farms are being developed around the world. In 2017, for example, the total installed capacity reached around 539 gigawatts (GW) \cite{globalwind2019} and this is expected to increase tremendously in the coming decades (See figure \ref{data}).

Despite the attractive features and benefits, wind energy generation comes with challenges, which power system operators face during integration into the electricity gird. These challenges arise mainly due to the non-stationary, limited predictability, limited dispatchability, and non-storability of the wind. Wind energy is not fully dispatchable since it is impossible to increase the wind power upon request; only to reduce it. Moreover, wind can not be stored (like atoms and coal) for future use \cite{zhu2012short}. In power system operation, the main objective is to ensure the electric supply and power demand are balanced in the face of uncertainties and transmission network constraints
 without incurring too much cost.

Power system operators decide ahead of time the amount of power supply produced to meet the predicted load demand and reduce cost. During the integration of large-scale wind power into electric grid the operation cost becomes high and system stability and reliability is reduced since wind is intermittent and difficult to predict ahead of time. The failure to produce enough supply due to wind power generators producing less power than predicted can result in huge loss such as blackouts or expensive mechanisms, such as gas-fired plants, are used to maintain balance \cite{xie2010wind}. Therefore, accurate and robust wind speed forecasting methods are important to integrate wind power into power systems and minimize costs. This work focuses on short-term wind speed forecasting since it is more accurate and reliable than other time-scale forecasting methods such as long-term. It is also ideal for effective load dispatch planning and real-time grid operation \cite{chang2014literature}.

LSSVM is an efficient model that is used in regression applications including wind speed prediction. To greatly improve the performance of this model, two of its parameters, the regularization parameter and the kernel parameter, need careful tuning. In this work we propose a new variant of QPSO, known as QPSO with elitist breeding (EBQPSO), to select optimum values of these two hyper-parameters and use the model to forecast short term wind speed for wind farms. To the best of our knowledge, this is the first work to propose short-term wind forecasting using LSSVM optimized by EBQPSO.

\section{Literature Review}
Accurate wind speed or wind power forecasting models are still in high demand. In recent years, scholars have done plentiful research on wind power prediction. Time series and AI-based algorithms methods are the most commonly used schemes. In time-series modeling, Autoregressive Moving Average (ARMA) methods are the most widely used techniques \cite{erdem2011arma}, \cite{liu2011comprehensive}, \cite{li2011wind}. 

Erdem \cite{erdem2011arma} employed four approaches based on ARMA methods to perform wind speed forecasting and direction tuple. After decomposing wind speed into lateral and longitudinal spaces, each component was forecasted using ARMA model and indvidual outputs were merged to provide final forecasts of wind speed and direction. Jiang \cite{jiang2012wind} a wind speed forecasting model using the hybrid of ARMA and generalized autoregressive conditional heteroscedasticity. The parameters of the proposed model were computed using quasi-maximum liklihood estimator with modified particle swarm optimization solving the solution of the hybrid model through log-quasi-likelihood function.

AI-based alogirthms approach wind forecast modeling by fitting historical wind data on a high dimensional non-linear function: examples include artificial neural networks networks (ANN) \cite{chang2013rbf}, \cite{wu2015short}, \cite{liu2012wind}, \cite{palomares2009arima} and LSSVM \cite{xiang2019forecasting}, \cite{sun2015wind}, \cite{hu2014short}. Chang developed a radial basis function (RBF) neural network by combining orthogonal least squares (OLS) and genetic algorithm and used the model for short-term wind speed forecasting in Taichung coast of Taiwan. The number of nodes was computed using the OLS algorithm while the parameters were selected using the genetic algorithm.

Wu et al. \cite{wu2015short} constructed a PSO-optimized B-spline neural network (BSNN) for shor-term wind speed forecasting. BSNN was implemented to generate input space variable and PSO was adopted to select number of internal nodes and prevent the easy fall of output into the local minimum. Compared to traditional BSNN, the proposed model achieved better perfromance. Khodayar et. al.\cite{khodayar2017rough} implemented deep neural networks for ultr-short and short-term wind speed forecasting in combination with  stacked auto-encoder (SAE) and stacked denoising auto-encoder (SDAE) for automated feature engineering from historical data. The proposed model proved to have better generalization capability and performance to shallow artificial neural netowrks.

The application of SVM for wind speed forecasting is discussed in some papers. Sun et. al. \cite{sreelakshmi2008performance} compared SVM with ANN for short-term wind speed forecasting. The prediction accuracy of SVM was found to be better than ANN with SVM displaying faster compuational time. SVM was also compared with a multilayer perceptron (MLP) neural network for wind speed prediction in \cite{mohandes2004support} in which the SVM model provded favorable results in terms of mean squared error. The Least-Squares Support Vector Machines (LSSVM) algorithm, which is a another version of SVM that solves linear equations rather than quadratic programming problems is also widely used in wind speed prediction. The performance of LSSVM is greatly dependent on the regularization parameter ($\gamma$) and kernel parameter ($\sigma ^ 2$). Inadvertent choice of these parameters can result in forecasting models that suffer from overfitting or underfitting. To circumvent this problem, researchers have proposed hybrid optimization algorithms to maintain better models.

Sun \cite{sun2012short} used PSO to generate optimum LSSVM parameters by training and validating the model on wind farm dataset collected from Inner Mongolia. Chang \cite{chang2013short} proposed a short-term wind speed forecasting model based on an enhanced PSO algorithm being employed for optimizing persistence method, neural network with back propagation, and radial basis function (RBF). Good agreements between the realistic values
and forecasting values were obtained. Since PSO algorithms have the tendency to fall in local minima while selecting optimum parameters of LSSVM, the generalization capability and learning performance can be affected. To overcome these problems, Hu \cite{hu2014short} introduced variant of QPSO that assigns more weight to particles with higher fitnesses and optimized parameters of LSSVM. and improved performance by changing. Although the proposed method significantly improved performance, the prediction accuray would further be improved if the weight coefficient was changed intelligently instead of linearly from 1.5 to 0.5.

\section{Least squares support vector machines}
LSSVM is a novel SVM that adopts least squares to recognize patterns in data for classication and regression problems. It was first proposed by Suykens as an extension of SVM to formuate a solution of minimization problem as linear rather than quadratic \cite{suykens1999least}. SVM is supervised machine learning algorithm with many applications in classification, regression, and function estimation problems. The algorithm converts optimization problems to quadratic programming with linear constraints and learns complicated patterns using functions known as kenels and also takes advantage of the sparseness of solutions. However, when the training size becomes larger the computational complexity also increases as the size of the quadratic programming is proportionally affected by it \cite{zhou2011fine}. LSSVM, on the other hand, 
converts inequality constraints into equality constraints, thereby eliminating some ambigous parameters, which also fascilitate for converting the quadratic programming into linear equations. This, in turn, produces robust loss function, speeds up convergence, and represses complexity of the solution \cite{hu2014short}.

The LSSVM algorithm works as follows: let a training data is represented as $\{(x_i, y_i), \, x_i \in \mathbb{R} ^ m, y_i \in \mathbb{R}, \, i = 1, \dots, N \}$, where $N$ is the size of the training point, and $m$ is the number of the feature inputs. A linear function is then used to fit the training set based on equation \eqref{linfun}.

\begin{equation}
\label{linfun}
f(x) = w^T \phi(x) + b
\end{equation}
where, $\phi(\cdot)$ is a nonlinear mapping function; $w$ is weight vector; $b$ is bias.

Following the risk minimization principle, we can transform equation \eqref{linfun} into a constrained minimization optimization problem as follows \cite{de2002least}: 

\begin{align}
\begin{split}
\label{optform}
\min J(w, \xi) = \frac{1}{2}w^Tw + \frac{1}{2}\gamma \sum_{i=1}^N \xi ^2 \\
\mbox{subject to } y_i = w^T\phi(x_i) + b + \xi _i,  \,\, i \in (1, \dots, N)
\end{split}
\end{align}

where, $\gamma$ is the error penalty parameter; $\xi_i$ is the slack variable. Using Largrange function the constrained optimization problem is transformed into an unconstrained optimization problem as follows:

\begin{equation}
\label{lagrang}
L = \frac{1}{2}w^Tw + \frac{\gamma}{2} \sum_{i=1}^N \xi^2 - \sum_{i=1}^N a_i(w^T\phi(x_i) + b + \xi _i - y_i)
\end{equation}

$a_i$ is Lagrange multiplier. After applying the Karush-Kuhn-Tucker (KKT) condistions, we obtain the following linear equation.
\begin{align}
\begin{split}
\label{kkt}
\left( \begin{array}{cc} 
		0 & \quad \textbf{I}^T \\
		1 & \quad \Omega+\gamma^{-1}\textbf{I} 
	\end{array} 
\right)
\left( \begin{array}{c} 
			b \\ a 
		\end{array} 
\right)
=
\left( \begin{array}{c} 
			0 \\ y
		\end{array} 
\right)
\end{split}
\end{align}

Where $y = [y_1, \dots, y_N]^T, \, a = [a_1, \dots, a_N]^T, \, \Omega = \phi(x_i)^T\phi(x_j), \, \mbox{and} \textbf{I}$ is a unit matrix.

Mercer condition establishes the kernel function as follows:

\begin{equation}
\label{kernel}
k(x_i, x_j) = \phi(x_i)^T\phi(x_j)
\end{equation}
We can derive $a$ and $b$ from equations \eqref{kkt} and \eqref{kernel} and formulate the final LSSVM regression function using equation \eqref{lssvm}.
\begin{equation}
\label{lssvm}
f(x) = \sum_{i=1}^N a_ik(x, x_i) + b
\end{equation}
For this work, we selected gaussian RBF kernel function since it is easy to implement and maps the training set into an infinite dimensional space making it ideal for problems with non-linear relationships. Equation \eqref{rbf} gives the expression for the guassian RBF function:
\begin{equation}
\label{rbf}
K(x, x_i) = \exp \big(\frac{-(x - x_i) ^ 2}{2\sigma^2} \big)
\end{equation}
here $\sigma$ is the width of core also called the kernel parameter.

The error penalty parameter $\gamma$ and the width of core $\sigma$ are important paramters of LSSVM possessing great impact on the overall performance of the algorithm. To obtain optimum values for these parameters we implemented the EBQPSO algorithm.

\section{QPSO with elitist breeding}
QPSO is a new variant of particles swarm optimization (PSO) inspired from quantum mechanics and probabilistic analysis of PSO. The algorithm updates its positions around the previous best points using the delta potential model \cite{sun2011quantum} and enhances the global search capabilities of its particles by exploiting the mean best position \cite{sun2004global}. The particles of QPSO are assumed to move in quantum space with no spin and the appearance of a particle at some position is driven from a probability density function \cite{zhou2008qpso}. With these proporties, the entire feasible space is searched to obtain optimal solutions. Given $M$ particles, each of which are represented in $n$ dimensional space, the position of particle $i$ during iteration $t$ is given as $X_i(t) = (X_{i, 1}(t), \dots, X_{i, n}(t))$ with its evolution governed using the following three equations:

\begin{equation}
\label{mbest}
m_{best}(t) = \frac{1}{M}\sum_{i=1}^M pbest_{i}
\end{equation}

\begin{equation}
\label{pci}
P_{c_{i, j}}(t) = \phi_{i, j}(t)*pbest_{i, j}(t) + (1 - \phi_{i, j}(t)) * gbest_j(t)
\end{equation}

\begin{equation}
\label{xit}
X_{i, j}(t+1) = P_{c_{i, j}}(t) \pm a*|m_j - X_{i, j}(t)|\ln(\frac{1}{u})
\end{equation}

$pbest_i(t)$ is called the personal best position, which is the best position of particle $i$ from previous iterations (i.e. the one with best fitness value). $m_{best}$ is the mean of the personal best positions of all particles.
$gbest(t)$ is the global best position, which the best position of all particles. $P_{c_{i}}$ is a random position between $pbest_i(t)$ and $gbest(t)$, $\phi$ and $u$ are any random number between 0 and 1 inclusive; $\alpha$ is the contraction expansion (CE) coefficient: it can be varied to adjust the convergence speed of the algorithm. Its value during the $t^{th}$ iteration is computed as follows:
\begin{equation}
\label{ce}
\alpha = 0.5 + 0.5 * (T - t)/T
\end{equation}
$T$ is the maximum number of iterations. 

\begin{algorithm}[!ht]
\label{alg:transposonop}
\caption{Procedure for transposon operator}
\begin{algorithmic}[1]
\Procedure{TransposonOperator}{$epool$}
	\LineComment{$epool$ is pool of particles}
	\State Define population size $M$, number of transposon $L$, and jumping rate $jrate$
	\State Generate $epool\_norm$ from $epool$ based on equation \eqref{norm}
	\For{$i = 1$ to $M+1$}
		\If{random(0, 1) $<$ $jrate$}
			\State $C_1 = i$
			\State $C_2 = ceil(rand(0,1) \times (M+1))$
		
			\If{$C_1 == C_2$}
				\If{random(0, 1) $>$ 0.5}
					\State Apply cut and paste operation in $epool\_norm[C1]$
				\Else
					\State Apply copy and paste operation in $epool\_norm[C1]$
				\EndIf
			\Else
				\If{random(0, 1) $>$ 0.5}
					\State Apply cut and paste operation in $epool\_norm[C1]$ 							\State and $epool\_norm[C2]$
				\Else
					\State Apply copy and paste operation in $epool\_norm[C1]$
					\State and $epool\_norm[C2]$
				\EndIf
			\EndIf
		\EndIf
	\EndFor
\EndProcedure

\end{algorithmic}
\end{algorithm}

In a regural QPSO the personal best of each particle and the global best of the entire population are simply stored for updating particle positions and comparing solutions despite being elitists of the algorithm. If explored and manipulated with care this elitist memory can contribute significant improvement in performance. In this paper, we propose QPSO with elitist breeding (EBQPSO) to search optimum parameters of LSSVM. The elitist breeding scheme forms a new subswarm through breeding of the elitists in the process of the evolutionary algorithm. In the process, an elitist pool denoted as $epool$, which consists of the most recent personal best particles and global best particle, is constructed. Subsequently, the transposon operator, which has the ability to enhance the diversity of solutions, is selected as the breeding operator.

The elitist breeding mimics the mutation of biological DNA elements known as transposons. Transposons are consecutive genes with their positions assigned randomly in chromosomes. Transposon operators are opertions that occur in one chromosome or between different ones in which genes are moved from one position to another. There are two types of transposon operators: cut and paste operators and copy and paste \cite{yang2015improved}. The choice of applying the operation is done randomly and the size of each transposon can be greater than or equal to one. A parameter called jumping percentage is used to choose the size of transposon. Also, the probability of activating the transposon operator is decided by a parameter called jumping rate.

If the jumping rate is greater than a randomly generated number between zero and one, transposon operation is carried out between the current chromosome and some randomly chosen chromosome. If the position of the current chromosome and the position of the randomly chosen chromosome are the same, this means the transposon operation occurs on the same chromosome. And the operation being cut and paste or copy and paste is randomly chosen. However, if the position of the current chromosome and a randomly chosen chromosome are not the same, the transposon operation should occur on different chromosomes. Again, the choice of cut and paste or copy and paste operation is random. The procedure for the transposon operator is shown in algorithm 1.

\begin{algorithm}[!ht]
\label{alg:ebqpso}
\caption{Procedure EBQPSO}
\begin{algorithmic}[1]
\State Define search space, population and fitness function $f$
\State Set population size $M$, dimension $n$, max iteration $T$
\State Generate random particles $\textbf{X}$
\State Initialize $pbest[i] = X[i] \quad 1 \leq i \leq M$
\For{$t = 1$ to $T$}
	\State $gbest = \argmin(f(pbest))$
	\State Compute $m_{best}$ based on equation 7.
	\If{elitist criterion met}
		\State $i = 1$
		\While{$i \leq M$}
			\State $epool[i] = pbest[i]$
			\State $i = i+1$
		\EndWhile
		\State $epool[M+1] = gbest$
		\State $epool\_eb=TransposonOperator(epool)$
		\For{i=1 to M}
			\If{$f(epool\_eb[i]) < f(pbest[i])$}
				\State $pbest[i] = epool\_eb[i]$
			\EndIf
		\EndFor
		\State $gbest = argmin(f(pbest)$
	\EndIf
	\For{i=1 to M}
		\State Compute $P_{c_i}$ based on equation \eqref{pci}.
		\State Update $X[i]$ based on equation \eqref{xit}.
		\If{($f(X[i])<f(pbest[i])$}
			\State $pbest[i] =X[i]$
		\EndIf
	\EndFor
\EndFor
\end{algorithmic}
\end{algorithm}

Each particle in QPSO can be regarded as a chromosome storing the same number of genes as the dimension of the particle, and each gene stores a real number corresponding to the dimensional value of a particle. Hence, the number of chromosomes is the same as the number of particles in the swarm. Since different dimnesional values in each particle might have different search spaces, the
dimensional values in each is normalized as follows:

\begin{equation}
\label{norm}
\textbf{x}_{norm} = \frac{\textbf{x} - max(\textbf{x})}{max(\textbf{x}) - min(\textbf{x})}
\end{equation}

where $min(\textbf{x})$ and $max(\textbf{x})$ represent the lower and upper bounds of $\textbf{x}$, respectively. Aftertransposon operation is carried out, the value of positional vector is restored back to itscorresponding positional value in the search space according to the following equation:

\begin{equation}
\label{denorm}
\textbf{x} = \textbf{x}_{norm} \times (max(\textbf{x}) - min(\textbf{x})) + min(\textbf{x})
\end{equation}

When the criteria are met, Algorithm 1 outputs a new subswarm $epool\_eb$ after applying the transposon operator is executed on the pool of the personal best of each particles and the global best collectively stored in $epool$. The $pbest$ is updated when the fitness value in the corresponding $epool$ is better than the newly generated individual. The frequency of the elitist breeding iscontrolled using a predefined parameter called lambda ($\lambda$). In every $\lambda$ iteration, the breeding operation will be performed once.

\begin{figure}[!ht]
\centering
\includegraphics[scale=.45]{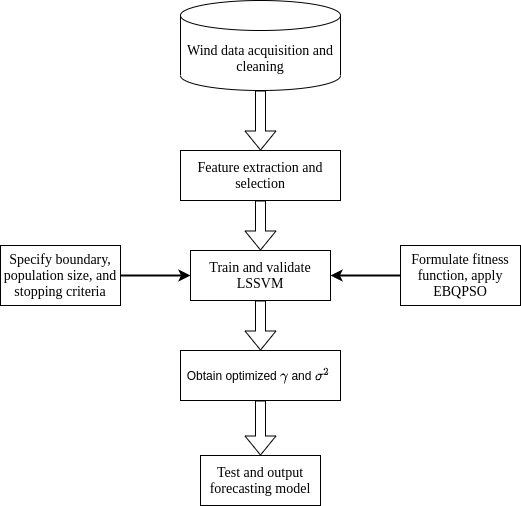}
\caption{\label{method} The overall methodlogy used.}
\end{figure}

\section{Methodology and anlysis procedure}
\subsection{Data acquisition and cleaning}
The wind speed dataset along with other parameters in a wind farm is collected from the supervised control and data acquisition (SCADA) system information. The dataset that is used for training and testing our model is obtained from Ashegoda wind farm, located in south of Mekelle city, Ethiopia. The data was collected from April 1, 2015 to May 31, 2015 constituting a total of 4,393 sample points. Each point is sampled every 20 minutes. The SCADA system does not always generate accurate data. Sometimes missing data points and outliers can be generated which can negatively affect the learning model and need to be handled well. In our case, the number of outliers and missing data points was very small compared to the size of the dataset. Therefore, in the preprocessing stage, all missing data and outliers are replaced with the mean of the wind speed dataset.

\subsection{Feature extraction and selection}
\begin{figure}[!ht]
\centering
\includegraphics[scale=.6]{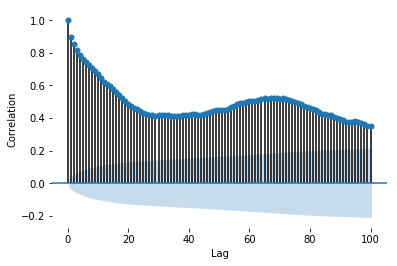}
\caption{\label{corr} Correlation values between lagged inputs and target output.}
\end{figure}

In this stage the given dataset is transformed into a new set of features. The wind speed dataset is a time series collection of univariate data points recorded every 20 minutes represented as a sequence $\{x_1, x_2, \dots\}$. From this, a sequence of data points $D := ((x_1, y_1), \dots , (x_N, \dots, y_N)$ of input/output pairs is constructed. We have $x \in \mathbb{R}^n$ and $y \in \mathbb{R}$ 
where $n$ is the number features to be selected. Wind speed can be effectively forecasted using its own historical data with LSSVM. We selected 100 historical data as feature inputs to our forecasting model, i.e, to predict a target value at time $t$, we obtained lagged data points at $(t-1, t-2, \dots, t-100)$ as input features. The motivation for selecting these features is due to the high correlation between these lagged values and the output value at time $t$ as depicted in Figure \ref{corr}.

The selection of feature inputs based on correlation values is not always a good approach since it may cause overfitting and increase training time due to high-dimensionality. Therefore, we selected only the best 10\% of the input features  using mutual information, which measure the degree of dependence between an input feature and the output data \cite{kraskov2004estimating}.

\subsection{EBQPSO based LSSVM parameters optimization}
\begin{figure}[!ht]
\centering
\includegraphics[scale=.45]{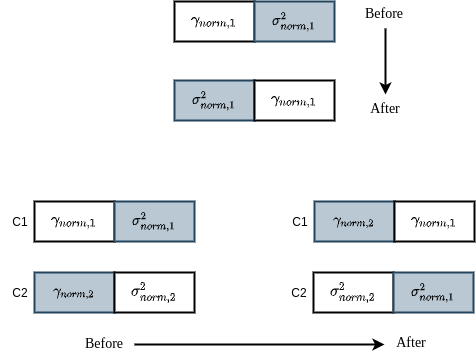}
\caption{\label{cutpaste} Applying cut-and-paste transposon operator on LSSVM parameters: (a) Same particle. (b) Different particles}
\end{figure}

\begin{figure}[!ht]
\centering
\includegraphics[scale=.45]{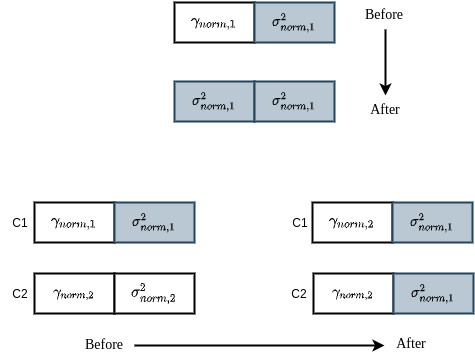}
\caption{\label{copypaste} Applying copy-and-paste transposon operator on LSSVM parameters (a) Same particle. (b) Different particles.}
\end{figure}

In order to generate optimum parameters of LSSVM using QPSO with elitist breeding (EB-QPSO), it is important to identify the parameters and how they can be represented as chromosomes. We have two parameters, the regularization parameter ($\gamma$), and the kernel parameter ($\sigma ^2$). Hence, each particle contains two positional values, which implies each chromosome will have two genes. This number will restrict the number of transposons to be one and the size of each transposon to be one as well. The cut-and-paste and copy-andpaste operations can be summarized as in Figure \ref{cutpaste} and Figure \ref{copypaste}.

The quality of the fitness function is a key measure of the evolutionary
algorithm. In this paper, the inverse of the root mean square error (RMSE), given in equation \ref{rmse}, is selected as the fitness function of the EB-QPSO algorithm.
\begin{equation}
\label{rmse}
f_{rmse}(\gamma, \sigma^2) = RMSE(\gamma, \sigma^2)
\end{equation}

Therefore, the objective of the EBQPSO-LSSVM algorithm will be to generate the regularization and kernel parameters of LSSVM that minimize the fitness function (i.e, the root mean squared error). The basic steps involved in optimizing the parameters are given as follows:
\begin{itemize}
\item[] \textbf{Step 1:} Determine the maximum number of iterations and the particle population size with dimension two as $X_i = (\gamma_i, \sigma_i^2)$. Choose the value of lambda.
\item[] \textbf{Step 2:} Use the training sample to train your model, the validation sample to obtain the fitness value of each particle according to equation 12.
\item[] \textbf{Step 3:} According to the calculated fitness value, update the optimal value $pbest$ of each particle and the global optimal $gbest$ of each particle. Compute $m_{best}$ according to equation 8a.
\item[] \textbf{Step 4:} If lambda is a factor of the current number of iteration, concatenate the $pbest$ and $gbest$ into one pool, $epool$. Perform transposon operation on $epool$. Update the $pbest$ and $gbest$ particles using the fitness of the new bred individuals. Otherwise, go to step 5.
\item[] \textbf{Step 5:} Update particle information according to equations 8b and 8c.
\item[] \textbf{Step 6:} Check whether the optimization condition is met. If maximum number of iteration is reached, end the optimization and output the current optimal particle position $(\gamma_i, \sigma_i^2)$. Otherwise, return to step 2
\item[] \textbf{Step 7:}  Use $(\gamma_i, \sigma_i^2)$ obtained through Step 6 to retrain and test LSSVM on unseen,  develop the regression model and report forecasting performance.

\end{itemize}

\section{Results and discussion}

\subsection{Experimental setup}
Table \ref{exp} presents the parameter settings for the EBQPSO algorithm. We set the maximum number of generations to 50 and the population size to 20. Since we are optimizing the two hyperparameters the problem dimension is set to 2, jumping percentage to 1, and number of transposons also to 1. The jumping rate is chosen to be 0.2 indicating the transposon operator is activated with 0.20 probability; otherwise, the algorithm continues with regular QPSO. The $\lambda$ value is set to 3 to initiate elitist breeding in every three generations. Selecting appropriate search space is also crucial: we set the minimum value of $\gamma$ to be 0.0001 and the maximum to be and minimum value of $\sigma^2$ to be 1 and the maximum value to be 4 $\times$ $10^4$. Also, we mapped the search space to log-scale to improve search capability and converge to optimum values with smaller number of iterations. 

\begin{table}[!ht]
\centering
\caption{\label{exp} Parameter setting for EBQPSO algorithm}
\begin{tabular}{|l|c|}
\hline
\textbf{Parameter} & \textbf{Parameter value} \\\hline
Number of generations (G) & 50 \\\hline
Population size (M) & 20 \\\hline
Problem dimension (d) & 2 \\\hline
CE coefficient ($\alpha$) for EB-QPSO & $\alpha = 0.5$ \\\hline
Jumping percentage & 1 \\\hline
Jumping rate & 0.2 \\\hline
Number of transposons & 1 \\\hline
Lambda ($\lambda$) & 3 \\\hline
Minimum search point & (0.0001, 8) \\\hline
Maximum search point & (1 $\times$ $10^6$, 4 $\times$ $10^4$) \\\hline
\end{tabular}
\end{table}

Finally, we used 60\% of our dataset for training, 20\% for validation (to compute and compare fitness values), and 20\% for testing our approach. The EBQPSO approach is compared with other competetive methods such as PSO and QPSO. Many standards for evaluating performance of prediction model are known. In this work, we use the mean absolute error (MAE), the root mean square error (RMSE), and the mean absolute percentage error (MAPE) given as follows.

\begin{equation}
\label{mae}
MAE = \frac{1}{N}\sum_{i=1}^N |y_i - \hat{y}_i|
\end{equation}

\begin{equation}
\label{rmse}
RMSE = \sqrt{\frac{1}{N}\sum_{i=1}^N |y_i - \hat{y}_i|^2} 
\end{equation}

\begin{equation}
\label{mape}
MAPE = \frac{1}{N}\sum_{i=1}^N \Big|\frac{y_i - \hat{y}_i}{y_i}\Big| \times 100
\end{equation}
 
\begin{figure}[!ht]
\centering
\includegraphics[scale=.45]{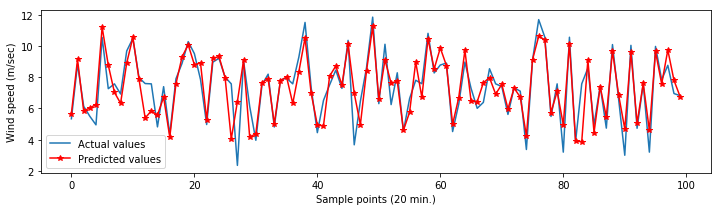}
\caption{\label{pred} Forecasted versus actual wind speed.}
\end{figure}

\begin{figure}[!ht]
\centering
\includegraphics[scale=.45]{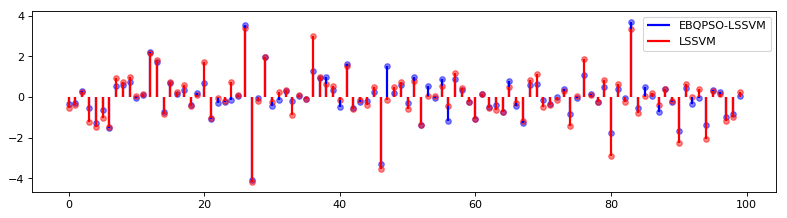}
\caption{\label{errors} Error distribution of proposed model compared with LSSVM.}
\end{figure}

\subsection{Experimental results and analysis}
Figure \ref{pred} shows the distribution of the forecasted wind speed versus the actual wind of our proposed model for 100 point samples from test set. It can be generalized that the model has performed well in predicting the samples of wind speed. The proposed model displays good prediction and generalization capabilities and detects most of the patterns generated in the observed wind speed. Further, we compared the forecasting error distribution of LSSVM optimized by elitist QPSO and LSSVM without optimization as shown in Figure \ref{errors} using 100 samples from the test set. It can be observed that the
EBQPSO-LSSVM model has lower errors in most of the samples. The
error distribution for EBQPSO-LSSVM model is invisible throughout most of the samples sheltered by the error distribution of LSSVM without optimization.

\begin{table}[!ht]
\centering
\caption{\label{perfor} Performance comparison of the three approaches using 5 trials given as \textbf{\textit{mean $\pm$ std}}.}
\begin{tabular}{|l|l|l|l|}
\hline
LSSVM optimizer strategy & RMSE & MAE & MAPE (\%) \\\hline
PSO & 0.992 $\pm$ 0.0408 & 0.699 $\pm$ 0.013 &  10.72 $\pm$ 1.08 \\\hline
QPSO & 1.046 $\pm$ 0.053 & 0.681 $\pm$ 0.037 & 9.88 $\pm$ 0.98  \\\hline
EBQPSO & \textbf{0.981 $\pm$ 0.061} & \textbf{0.671 $\pm$ 0.051} & \textbf{9.43 $\pm$ 1.12} \\\hline
\end{tabular}
\end{table}

Table \ref{perfor} also compares the performance of the proposed approach with PSO and QPSO approaches using five experimental trials for each methods and using the mean and standard deviation as a final evaluation metric. As we can see from the table the PSO and QPSO achieved a similar performance in most cases with PSO scoring the worst mean average percetange error of 10.72\%. Also, the QPSO approach achieved a better mean average error (6.81\%) compared to the PSO approach while the PSO performed better in terms of root mean square error (0.992) compared to the QPSO. The proposed EBQPSO outperformed the two methods in terms of all three evaluation metrics. We believe that this performance difference can be increased if more trials are used.

\section{Conclusion}
At present, renewable energy and machine learning are the fastest growing fields of study due to the alarming need to protect environment and energy, and provide cost efficient and automated mechanisms to do so. In this work we proposed an improved optimization algorithm to tune parameters of LSSVM, thereby increase the performance of the prediction model which is trained using a dataset obtained from a local wind farm. The LSSVM model optimized using the new variant of QPSO, known as QPSO with elitist breeding, has shown a  better forecasting capability compared to traditionla LSSVM. Furthermore, the proposed model achieved a better performance in terms of root mean square error, mean average error, and mean average percentage error compared to PSO and QPSO approaches.

\bibliographystyle{unsrt}  
\bibliography{references}  


\end{document}